\documentclass[final,1p,times]{elsarticle}

\usepackage{amssymb}
\usepackage{slashed}
\usepackage{lipsum}
\usepackage{amsfonts}
\usepackage{graphicx}
\usepackage{epstopdf}
\usepackage{algorithmic}
\usepackage{amsmath} 
\usepackage{soul}
\usepackage{color}
\usepackage{changes}
\usepackage{cases}
\usepackage{cleveref}
\usepackage{amsopn}
\usepackage{array}
\usepackage{booktabs} 
\usepackage{multirow}
\usepackage{booktabs}
 
\usepackage{threeparttable} 
\usepackage{url} 
\usepackage{calc}

\usepackage{multirow}
\usepackage{caption}
\usepackage{amsthm}
\usepackage[thicklines]{cancel}
 \usepackage{soul} 
 \usepackage{makecell}
 \usepackage{subfigure}

\newdefinition{remark}{Remark}
\newdefinition{definition}{Definition}

\allowdisplaybreaks[4]

\usepackage{lineno}

\begin{document}
\captionsetup[figure]{labelfont={bf},labelformat={default},labelsep=period,name={Fig.}}

\begin{frontmatter}



\title{Graph Attention Hamiltonian Neural Networks: A Lattice System Analysis Model Based on Structural Learning\tnoteref{t1}}

\author[address1]{Ru Geng\corref{cor2}}

\author[address1]{Yixian Gao\corref{cor2}}

\author[address1]{Jian Zu\corref{cor1}}
\ead{zuj100@nenu.edu.cn}

\author[address2,address3]{Hong-Kun Zhang\corref{cor1}}
\ead{hongkun@math.umass.edu}

\cortext[cor2]{These authors contributed equally to this work.}
\cortext[cor1]{Corresponding author}

\address[address1]{Center for Mathematics and Interdisciplinary Sciences, and School of Mathematics and Statistics, Northeast Normal University, Changchun, 130024, China}
\address[address2]{Department of Mathematics and Statistics, University of Massachusetts Amherst, MA 01003 USA}
\address[address3]{Department of Mathematics, Great Bay University, Dongguan, 523000, P.R. China}

\begin{abstract}
A deep understanding of the intricate interactions between particles within a system is a key approach to revealing the essential characteristics of the system, whether it is in-depth analysis of molecular properties in the field of chemistry or the design of new materials for specific performance requirements in materials science. To this end, we propose Graph Attention Hamiltonian Neural Network (GAHN), a neural network method that can understand the underlying structure of lattice Hamiltonian systems solely through the dynamic trajectories of particles. We can determine which particles in the system interact with each other, the proportion of interactions between different particles, and whether   the potential energy of interactions between particles exhibits even symmetry or not. The obtained structure helps the neural network model to continue predicting the trajectory of the system and further understand the dynamic properties of the system.
In addition to understanding the underlying structure of the system, it can be used for detecting lattice structural abnormalities, such as link defects, abnormal interactions, etc.  These insights benefit system optimization, design, and detection of aging or damage.
Moreover, this approach can integrate other components to deduce the link structure needed for specific parts, showcasing its scalability and potential. We tested it on a  challenging molecular dynamics dataset, and the results proved its ability to accurately infer molecular bond connectivity, highlighting its scientific research potential.
 \end{abstract}



\end{frontmatter}

\section{Introduction}\label{sec1}

The impact of lattice systems on scientific discovery and technological advancement is profound and far-reaching. These systems are pivotal in various fields, including condensed matter physics \cite{2006Ultracold}, materials science \cite{Curtarolo2012}, chemistry \cite{Coe2019}, biology \cite{Mark2004}, biochemistry \cite{Abkevic1994}, and medical research \cite{hu2019}.
Lattice systems have also received attention among mathematical physicists, such as the statistical mechanics of lattice systems \cite{Friedli2017Statistical} and nonlinear lattice theory \cite{Toda2012}.
The lattice structure reflects the interaction between particles. A deep understanding of lattice structures is invaluable across various scientific domains. In chemistry, for example, the exploration and synthesis of new compounds often hinge on knowledge of lattice structures. Materials scientists exploit this understanding to design novel materials with tailored properties by analyzing atomic or molecular arrangements and interactions; In addition, studying lattice defects is also an important topic of their research.
Similarly, in pharmacology, the physical structure and properties of drug solids guide the development of effective drug products \cite{glazer2016crystallography, oganov2019structure, datta2004crystal,pond1994defects}.

In this paper, we mainly study the interaction relationship of a class of coupled nonlinear lattice systems composed of $N$ point-like particles $\{\alpha_i\}_{i=1}^N$, which is governed by a function called Hamiltonian $H: \mathbb{R}^{2N} \rightarrow \mathbb{R}$.
 It is expressed as 
\begin{equation}\label{dpq}
\frac{d}{dt}\left(
\begin{array}{c} 
\mathbf{q}\\
\mathbf{p}
\end{array}
\right)=
J
\left(
\begin{array}{c} 
\frac{\partial H(\mathbf{q},\mathbf{p})}{\partial \mathbf{q}}\\
\frac{\partial H(\mathbf{q},\mathbf{p})}{\partial \mathbf{p}}
\end{array}
\right),\quad
J=
\left(
\begin{array}{c c} 
O & I\\
-I&O
\end{array}
\right).
\end{equation}
The vectors $\mathbf{q}=(q_1,\cdots,q_N) \in \mathbb{R}^N$ and $\mathbf{p}=(p_1,\cdots,p_N) \in \mathbb{R}^N$ are generalized position and generalized momentum vectors in the phase space $\mathbb{R}^{2N}$, respectively. 
$I\in \mathbb{R}^{N\times N}$ is the identity matrix, $O \in \mathbb{R}^{N\times N}$ is the zero matrix.
$H$ is the Hamiltonian function. Commonly, it can be represented as
$H=\sum_{i=1}^N F_i(q_{\mathcal{N}_k(i)},p_i)$, where  $\mathcal{N}_{k\in M}(i)$   denotes $k$-th order neighbors of particle $\alpha_i$, $ M \subset \{0,1,\cdots,N-1\}$.
The function $F_i$ is related to the kinetic energy and potential energy of particle interactions about $\alpha_i$.
 This model is applicable to various problems, including cosmology, biology, thermal conductivity, atomic vibrations in crystals and molecules, and field modes in optics or acoustics \cite{kevrekidis2020emerging,tekic2016ac,kevrekidis2019dynamical, macia2009charge}.
 
\begin{figure*}
\centering
\includegraphics[width=.85\linewidth]{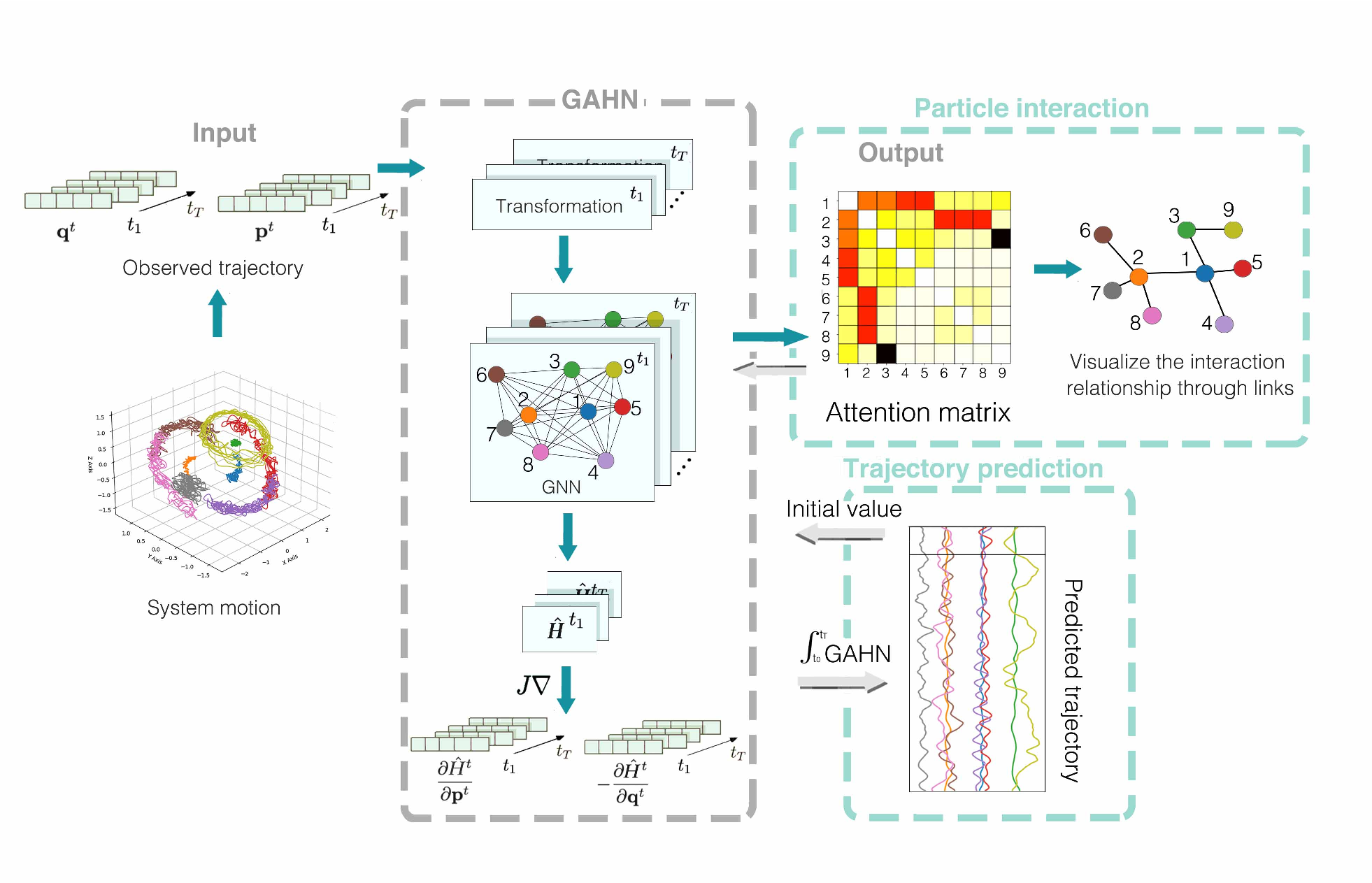}
\caption{The architecture of the GAHN.}
\label{SGHNfig}
\end{figure*}

The emergence of artificial neural networks represents a transformative development in the analysis and interpretation of complex data.  In the specific context of lattice systems, deep learning has enabled significant advances in delineating governing equations, identifying phase transformations, and crafting predictive models. However, these developments have primarily focused on simpler lattice systems, which are characterized by short-range interactions, often in one-dimensional constructs \cite{saqlain2022discovering, zvyagintseva2022machine, li2022gradient, zhu2022neural, jin2022learning}. This leaves a substantial gap in understanding and modelling the more complex lattice systems that feature long-range interactions and multi-dimensional structures, an area where traditional deep learning approaches have yet to make their mark.

There are currently some models that utilize Graph Neural Network (GNN) to learn lattice systems \cite{verdon2019quantum, kochkov2021learning, bishnoi2023learning,sanchez2019hamiltonian, geng2024separable}.
While these methods can improve the accuracy of system trajectory prediction, they require precise knowledge of the graph structure of the system, essentially the existence of interactions within the system. Unlike universal gravitation in celestial bodies \cite{bishnoi2023learning,sanchez2019hamiltonian}, which can be modelled using a fully connected graph due to its omnipresence, interactions in most lattice problems are complex and varied. This complexity makes it challenging to visually identify which particles interact, especially when dealing with hidden long-range interactions. Consequently, although graph neural network-based methods are effective, they necessitate more input information-specifically, the system's structure-compared to conventional neural networks.

To address these challenges, we propose a novel method, Graph Attention Hamiltonian Neural Networks (GAHN), which decodes the interaction potential among point-like particles from their dynamic trajectories by introducing a matrix, which we call the attention matrix and represent it as $\mathbf{A}=(a_{ij})$.
Unlike existing graph attention mechanisms such as Graph Attention Networks (GAT) \cite{velivckovic2017graph}  or Graph Convolutional Networks (GCN) \cite{kipf2016semi}, which rely on input data or the number of neighbors, our attention matrix learns the intrinsic properties of the system. By incorporating our defined graph learning loss \(\mathcal{L}_{GL}\), it accurately captures the structure of the system.

Our methodology begins with a graph representation of the Hamiltonian lattice system, where each particle \(\alpha_{i}\) (for \(i = 1, \cdots, N\)) corresponds to a node \(v_i\) on the graph. Given that the system structure is initially unknown, we start by initializing the graph as a fully connected directed graph with \(N\) nodes. The rows and columns of the matrix \(\mathbf{A}\) correspond to graph nodes (or system particles) and are indexed in the same order. The element \(a_{ij}\) encodes the relationship between nodes \(v_i\) and \(v_j\) (or particles \(\alpha_i\) and \(\alpha_j\)). Through this learned matrix, we can infer the structure of the lattice system, including which particles interact with each other, the proportion of interactions between different particles, whether the interaction potential energy between particles exhibits even symmetry, whether there are connection defects, whether there are impurity defects causing different interactions, and so on.
 In addition, we also demonstrate an extended application based on learning system interaction - using the obtained interaction information for trajectory prediction. This performs much better than conventional baseline models that do not consider interactions.

 The core framework of GAHN can be integrated into other tasks to learn the structure (interactions) of the studied system based on the task at hand. To illustrate this point, we validated our method's ability to learn chemical bond connections of molecules on a molecular dynamics dataset. The results obtained by GAHN are consistent with the existence of chemical bonds provided in current chemistry textbooks and research literature. This highlights GAHN's potential as a versatile tool for advancing the study of lattice systems through trajectory data.

Fig. \ref{SGHNfig} provides a visual representation of the GAHN architecture.
 
 The key contributions of our work are:

i) We propose a model capable of learning the underlying structure of  lattice Hamiltonian system solely through dynamic trajectories, without requiring any prior knowledge of the structure. This enables the method to learn lattice particle interaction potentials, particularly the hidden long-range potentials that are not intuitively observable. This has positive implications for understanding the structure of lattice systems and detecting anomalies in lattice structures, etc.

ii) Our method provides the underlying system structure for graph neural networks, addressing the limitation of graph neural networks require prior knowledge of system structure, unlike conventional neural networks.

iii) Our method can also integrate other components to infer the appropriate link structure required for specific applications. We validated it on a molecular dynamics dataset, and our method correctly inferred the connectivity of molecular chemical bonds. This indicates that our method has the potential to be widely applied in scientific work.

\section{Results}

\subsection{Hamiltonian lattice systems}
We choose three systems for our testing. The 1D Klein-Gordon lattice system with second-order and third-order long-range interactions (KG-LRI)\cite{penati2019nonexistence,koukouloyannis2013multibreathers,efremidis2002discrete}  is used to test the effectiveness of our model in capturing long-range interactions, and its potential energy is even symmetric, where we set $N=32$. 
The 2D Fermi-Pasta-Ulam-Tsingou (FPUT) system \cite{dauxois2005anti,pelinovsky2022kp,onorato2015route} is used to test the performance of our model on systems with high spatial dimensions and multiple interactions, and it contains odd potential energies, where we set $N=8\times 8$. The 1D Toda system \cite{ford1973integrability}  is used to test the effectiveness of exponential potential interactions, where we set $N=32$. See Table \ref{lattice_model} for their expressions. 
To demonstrate the effectiveness of model structure learning is independent of the arrangement of input data vector elements, We disrupt the particle order of the dataset about Toda.
 The order of the data vector elements we receive does not reflect the actual links of the chain, which are as follows $\alpha_{22}-\alpha_1-\alpha_{17}-\alpha_7-\alpha_{14}-\alpha_9-\alpha_{18}-\alpha_{19}-\alpha_{16}-\alpha_{27}-\alpha_{23}-\alpha_{2}-\alpha_{15}-\alpha_{30}-\alpha_{20}-\alpha_{26}-\alpha_{10}-\alpha_{5}-\alpha_{21}-\alpha_{11}-\alpha_{28}-\alpha_{32}-\alpha_{6}-\alpha_{31}-\alpha_{29}-\alpha_{12}-\alpha_{8}-\alpha_{3}-\alpha_{13}-\alpha_{25}-\alpha_{24}-\alpha_{4}$.
Without losing generality, the order of the indices of graph attention matrix $\mathbf{A}$ used in GAHN corresponds to the order of the vector elements 
in the obtained data, i.e. $\alpha_1,\alpha_2,\cdots,\alpha_{32}$.

To generate trajectory data, we employ the explicit symplectic Euler method over a $[0, 30]$ second interval with a time step of 0.001 and a stringent error tolerance of $10^{-12}$. Initial conditions (ICs) for 1D are set as:
\begin{eqnarray}
&&q^0_i=\lambda_{i} \sin\bigg(\frac{(i-1) \pi}{N-1}\bigg),\nonumber\\ 
&&p^0_i=0,\quad i=1,\cdots,32,
\end{eqnarray}
and for 2D are set as:
\begin{eqnarray}
&&{q}^0_{i,j}=\lambda_{i,j} \sin\bigg(\frac{\big(M(i-1)+(j-1)\big) \pi}{MN-1}\bigg),\nonumber\\ 
&&{p}^0_{i,j}=0,\quad i, j=1,\cdots,8.
\end{eqnarray}
with $\lambda_{i}$ and $\lambda_{i,j}$ drawn from a standard normal distribution.
Periodic boundary conditions are applied, and trajectories are sampled every 0.2 seconds to create the dataset, simulating high-precision observational equipment. The dataset consists of 70 trajectories and is partitioned into training and test sets in a 5:2 ratio.

The total number of epoch for all model training is set to 10000.
The optimizer is Adam, and the batch size is 256.
We adopt a learning rate piecewise constant decay strategy \cite{montavon2012neural}. 
The learning rate is initialized to $10^{-3}$, decays to $10^{-4}$
after 3500 epochs, and decays to $10^{-5}$
after 5000 epochs.

\subsubsection{Potential interaction learning}

\begin{table*}
\caption{Nonlinear lattice systems}\label{lattice_model}
  \centering
  \renewcommand\arraystretch{2.2}
  \begin{tabular}{|c|c|}
    \hline
      System &
 Hamiltonian\\
\hline
KG-LRI & $H=\sum_{i=1}^{32}\big(\frac{1}{2}p_i^2+\frac{1}{2}q_i^2+\frac{1}{4}q_i^4+\frac{1}{4}(q_i-q_{i+1})^2
 +\frac{3}{20}(q_i-q_{i+2})^2+\frac{1}{10}(q_i-q_{i+3})^2\big)$ \\
\hline
2D FPUT & $H=\sum_{i=1}^8 \sum_{j=1}^8\big(\frac{p_{i,j}^2}{2}+ V(q_{i+1,j}-q_{i,j})
+ V(q_{i,j+1}-q_{i,j})\big)$, with $V(u)=\frac{1}{2}u^2+\frac{1}{12}u^3$                                                                            \\ 
\hline
Toda          & $H=\sum_{i=1}^{32} \big(\frac{p_i^2}{2}+\exp(q_i-q_{i+1})\big)$                                     \\ 
\hline
  \end{tabular}
\end{table*}

The Attention matrix $\mathbf{A}=(a_{ij})$  reflects the interaction information of systems. The row and column order of $\mathbf{A}$ is the same, corresponding to particles. The specific definition of network model can be found in the supplementary materials.
 The magnitude of the $a_{ij}$  with $i\neq j$ encodes the strength of the interaction potential energy between particles  $\alpha_i$ and $\alpha_j$ (It can be understood as the 'elastic coefficient' of an invisible spring). And $a_{ii}$ encodes the energy generated by the particles themselves.
In addition, if $\mathbf{A}$ is a symmetric matrix, then the potential energy function of the system has even symmetry, and the interaction between particles $\alpha_i$, $\alpha_j$ is equivalent in both directions.

\begin{figure*}[htbp]%
\centering
\includegraphics[scale=0.75]{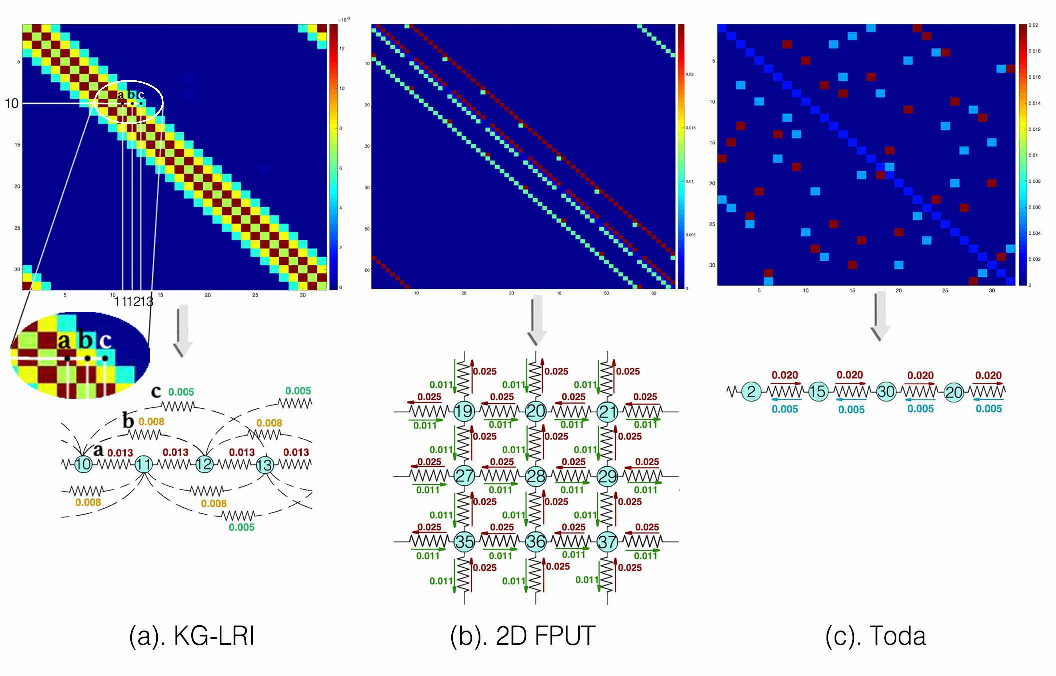}
\caption{The heatmap of the matrix $\mathbf{A}$ (top) and local graph structure inferred by the matrix (bottom).
For clarity, we provide a kind of particle space coordinate. The values on the edges of the structural graph correspond to the coefficients between two particles in the matrix. Since $a_{ij}\approx a_{ji}$ in $\mathbf{A}$ for KG-LRI, we only record values in one direction on the edges of the right graph and do not mark the direction of the edges.  }\label{fig_atten}
\end{figure*}

Fig. \ref{fig_atten} (a) demonstrates the effectiveness of our method in capturing long-range interactions. In Fig. \ref{fig_atten} (a) (top), the heatmap illustrates the attention matrix of the KG-LRI system, which has been learned through GAHN.
Based on this heatmap, we can deduce that the KG-LRI system exhibits three distinct types of interactions, indicated by the three different colors in the matrix elements $a_{ij}$ where $i<j$. The interaction between particles $\alpha_i$ and $\alpha_j$ is equivalent, as the potential energy of the system is even symmetric. Additionally, it appears that the dataset employs periodic boundary conditions.
In the heatmap, the dark blue regions represent values close to zero, indicating a lack of interaction between particles, while other colored areas depict varying intensities of interaction forces. Following the logic of the heatmap , whenever an interaction is identified between particles, we use springs to connect these particles, leading to the visualization presented in Fig. \ref{fig_atten} (a) (bottom).
Considering the 10th row of the matrix as an example, we observe that the values corresponding to columns 11, 12, and 13 are notably non-zero, as evident in the enlarged view marked by a, b, and c. Consequently, the 10th particle is connected to the 11th, 12th, and 13th particles on its right.
This analysis reveals that KG-LRI can be represented as a one-dimensional chain structure involving second-order and third-order long-range interactions.
 Alternatively, it can be modeled as a zigzag structure characterized by short-range interactions, which is analytically equivalent to a chain structure with long-range interactions see \cite{penati2019nonexistence}. The crucial aspect lies in the interactions among particles, rather than the specific spatial sequence of their coordinates.
Based on the coefficients of the matrix $\mathbf{A}$, we can infer the proportional relationship of the interaction potential between particles and their nearest neighbors of different orders. 
Evidently, the potential energy ratio between a particle and its first, second, and third nearest neighbors is roughly $5:3:2$. This finding aligns with coefficients of the test equation, namely $1/4:3/20:1/10$, which simplifies to the same ratio of $5:3:2$.
This conclusion is consistent with our initial testing assumptions.

Fig.  \ref{fig_atten} (b) validates the ability of our method to capture the interactions of systems with high spatial dimensions.
Fig. \ref{fig_atten} (b) (top) shows the heatmap of the attention matrix for the FPUT system.
We can see that each particle interacts with four other particles, with each of the four interactions being of equal proportion, and the test dataset employs periodic boundary conditions.
The potential energy of the system is asymmetric. In fact, the FPUT we tested incorporates a potential energy term of the third power, thus lacking even symmetry. The sum of potential energy interactions for each particle remains constant. 
Without considering long-range interactions, for clarity, we represent the particles as a $8\times8$ grid  and connect them based on $\mathbf{A}$. The structure of the local graph is shown in Fig. \ref{fig_atten} (b) (bottom).

Fig. \ref{fig_atten} (c)  verifies that the effectiveness of our method is independent of the arrangement of input data vector elements and demonstrates its effectiveness for exponential potential energy.
Fig. \ref{fig_atten} (c) (top) shows the heatmap of the attention matrix for the Toda system. For datasets with disordered particle order, the attention matrix learned by GAHN accurately represents the connections between particles within those datasets. 
The Toda model's exponential potential gives rise to asymmetric potential energy between particles.
However, the sum of potential energies between particles remains constant. Fig. \ref{fig_atten} (c) (bottom) shows the inferred local structure graph of the system.

\subsubsection{The role of $\mathcal{L}_{GL}$}
\begin{figure}%
\centering
\includegraphics[scale=0.3]{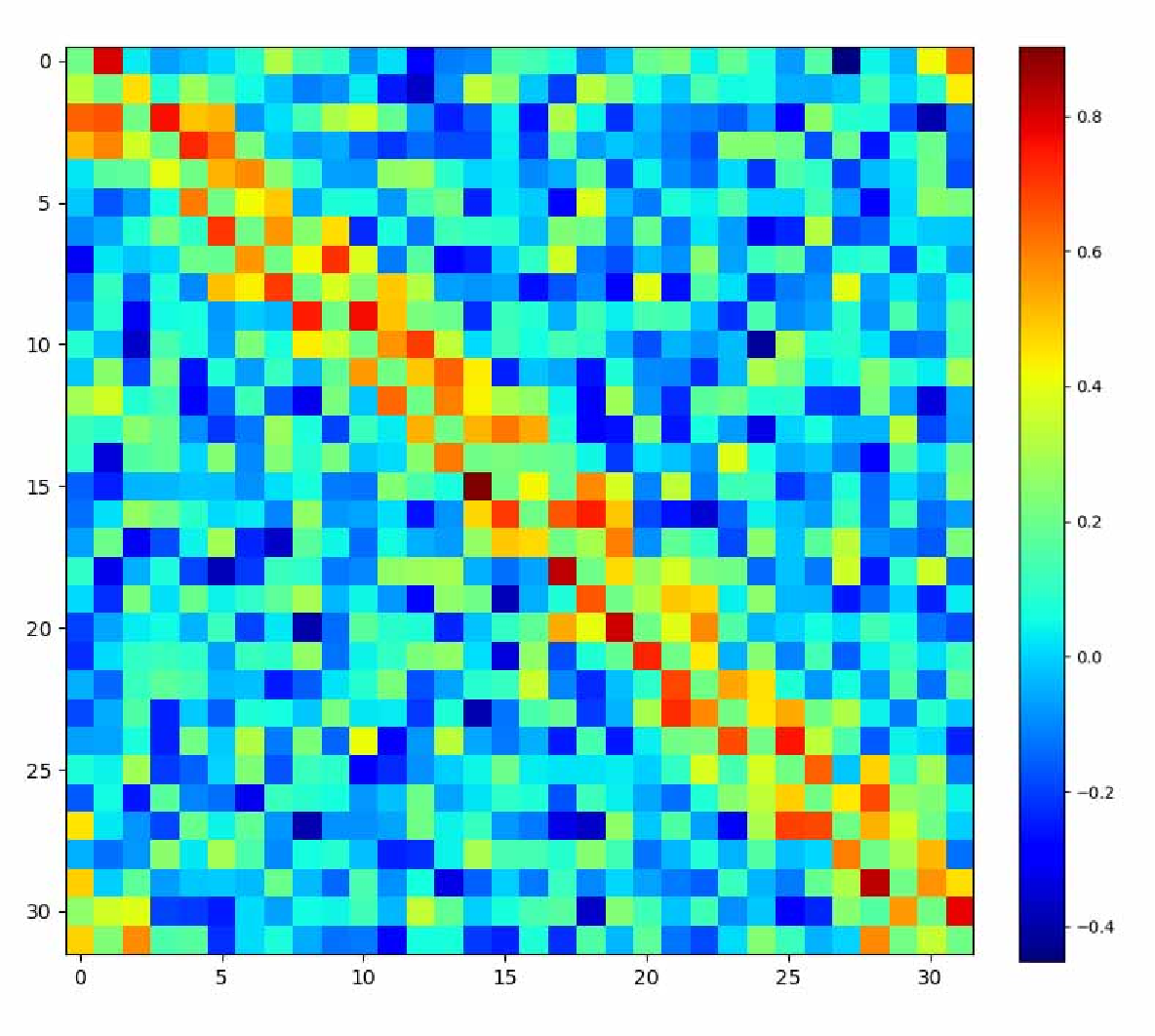}
\caption{The heatmap of the matrix $\mathbf{A}$ for KG-LRI when not considering the loss $\mathcal{L}_{GL}$.  }\label{fig_kglri}
\end{figure}

Fig. \ref{fig_kglri}  shows that without loss $\mathcal{L}_{GL}$, $\mathbf{A}$ will not be able to correctly capture the interaction relationship between particles.

 \subsubsection{System trajectory prediction} 

 In this section, we use the learned graph structure for further training in GAHN. Specifically, we process the learned attention matrix by applying a threshold of 0.001 (Note that choosing this threshold is straightforward because the weights of non-existent edges are negligible, and there is a notable disparity between the weights of existing edges. We can select any point within this disparity as the threshold). Values below the threshold are assigned zeros, and then input the attention matrix into the model for training. Note that at this stage, the parameters of the attention matrix remain fixed and are not involved in the training process. Our graph structure can also be utilized by other graph neural network models, enabling them to perform effectively.

To ascertain the effectiveness of our proposed model's prediction performance, we conducted experiments comparing GAHN with established models such as Multi-layer Perceptrons (MLP) \cite{popescu2009multilayer}, Hamiltonian neural networks (HNN) \cite{greydanus2019hamiltonian}, and Symplectic Networks (SympNet) \cite{jin2020sympnets} in both LA-type and G-type configurations. 
Given our assumption that only the trajectory dataset is known, existing system learning models based on graph neural networks are unsuitable as baselines. This is because they require prior knowledge of system interactions to construct the graph's edges based on these interactions.
The LA-SympNet was structured with a depth of 20 and incorporated four sublayers, while the G-SympNet was designed with a depth of 20 and a width of 50. MLP and HNN were each configured with three layers and 200 hidden units. SympNets employed sigmoid activation functions, aligning with the specifications detailed in Jin et al. \cite{jin2020sympnets}, whereas other models utilized SiLU functions.

To evaluate our model using the test set, we tracked two metrics: 

(i). Mean Squared Error (MSE) of the predicted energies ($H$). This metric assesses whether the network model adheres to the property of system energy conservation during long-term predictions. 

(ii). MSE of the predicted trajectories $(\hat{\mathbf{q}},\hat{\mathbf{p}})$. This metric gauges the stability of the network model in predicting generalized momentum and position over extended timespans

To evaluate the predicted energy and trajectories over extended timespans,
we integrated neural network models using the following equation:
\begin{eqnarray}
(\hat{\mathbf{q}}^t,\hat{\mathbf{p}}^t)=(\mathbf{q}^0,\mathbf{p}^0)+\int_{t_0}^t {\mathcal{N}_{\theta}} dt.
\end{eqnarray}
This integration was performed using a fourth-order Runge-Kutta integrator. Here, $(\mathbf{q}^0,\mathbf{p}^0)$ represents the initial values from the test set, while $(\hat{\mathbf{q}}^t,\hat{\mathbf{p}}^t)$ denotes the predicted generalized coordinates and momentum.
In our experiments, we set $t_0=0$, $t=30$, with a time step size of 0.002.
${\mathcal{N}}_{\theta}$ stands for various neural network models, including MLP, HNN and GAHN.
 For the SympNet model, the predicted trajectories were obtained using the equation:
\begin{eqnarray}
(\hat{\mathbf{q}}^{t+1},\hat{\mathbf{p}}^{t+1})={\mathcal{N}}_{SympNet}(\hat{\mathbf{q}}^t,\hat{\mathbf{p}}^t).
\end{eqnarray}
Note that the time step of SympNet should align with the training step of 0.2 to achieve good results. Here $\mathcal{N}_{SympNet}$ represents both LA type and G type SympNets.

The MSE of the predicted trajectories is defined as
\begin{eqnarray}
\rm{MSE}_{traj}=\sum_{\alpha \in \mathbb{Z}^n}\bigg((q_{\alpha,t}-{\hat{q}}_{\alpha,t})^2+(p_{\alpha,t}-{\hat{p}}_{\alpha,t})^2\bigg).
\end{eqnarray}

Table \ref{energy_error} and Table \ref{trajectories_error}
 respectively record the MSE of predicted energy and trajectory. The MSE between the ground truth and prediction trajectories from 20 prediction samples.  The best results are emphasized by bold fonts.  
Obviously, GAHN has performed exceptionally well.

\begin{table}
  \caption{The MSE of prediction energy.}
  \label{energy_error}
  \centering
    \resizebox{\columnwidth}{!}{
  \begin{tabular}{cccc}
   \toprule
    & KG-LRI    & 2D FPUT    & Toda\\
 \midrule
MLP       &    8.04E-3 $\pm$  1.60E-2  &   1.01E+1 $\pm$  3.55E+0      &   8.15E-4 $\pm$  9.28E-4      \\
HNN    &  2.17E-5 $\pm$  2.45E-5   &   6.75E-4 $\pm$  1.12E-3  &   4.66E-6   $\pm$  1.08E-5  \\
LA-SympNet    &  7.24E-2  $\pm$  3.41E-2  &  3.03E+1   $\pm$1.32E+2    & 2.92E-3   $\pm$  4.18E-3   \\
G-SympNet      &  7.50E-2  $\pm$  4.08E-2  &  4.55E-1  $\pm$ 1.98E+0    & 3.00E-3  $\pm$ 4.60E-3  \\
GAHN(ours)  &  \textbf{1.14E-7 $\pm$ 1.62E-7}   &\textbf{ 8.12E-8 $\pm$ 9.62E-8 }   & \textbf{  1.06E-7 $\pm$1.24E-7 }   \\
   \bottomrule
  \end{tabular} }
\end{table}

\begin{table}

  \caption{The MSE of prediction trajectories.}
  \label{trajectories_error}
  \centering
  \begin{threeparttable} 
  \resizebox{\columnwidth}{!}{
  \begin{tabular}{cccc}
 \toprule
    & KG-LRI    & 2D FPUT    & Toda\\
 \midrule
MLP       &     6.48E-3 $\pm$  1.24E-2   &    6.64E-3  $\pm$ 7.99E-3     & 2.20E-3   $\pm$  3.36E-3      \\
HNN    &  2.71E-3  $\pm$  5.79E-3  &  1.16E-2   $\pm$ 1.42E-2  & 1.27E-4    $\pm$  4.51E-4     \\
LA-SympNet    &  7.80E-2  $\pm$  7.09E-2  &  1.67E-1   $\pm$ 4.10E-1   & 6.44E-2    $\pm$  8.16E-2    \\
G-SympNet      &    7.23E-2  $\pm$  8.71E-2   & 9.70E-3   $\pm$7.60E-3     & 1.42E-2  $\pm$1.24E-2  \\
GAHN(ours)  &  \textbf{7.45E-7 $\pm$ 7.59E-7}   &\textbf{ 1.34E-6 $\pm$ 2.29E-6 }   & \textbf{  1.38E-6 $\pm$1.57E-6 }   \\
  \bottomrule
  \end{tabular} }
\end{threeparttable}
\end{table}

 \subsubsection{Lattice structure anomaly detection} 

Another function of GAHN is to detect anomalies in the lattice structure by observing trajectory data, such as connection fracture defects or impurity defects that cause different interactions.
Taking KG-LRI as an example, let's assume three scenarios occur in this system: the interaction between the third and fourth particle is 0; the long-range interaction coefficient between the third and sixth particle is twice as high as before; there is an additional interaction between the tenth and twentieth particle with a coefficient of 0.7.

\begin{figure}[htbp]%
\centering
\includegraphics[scale=0.8]{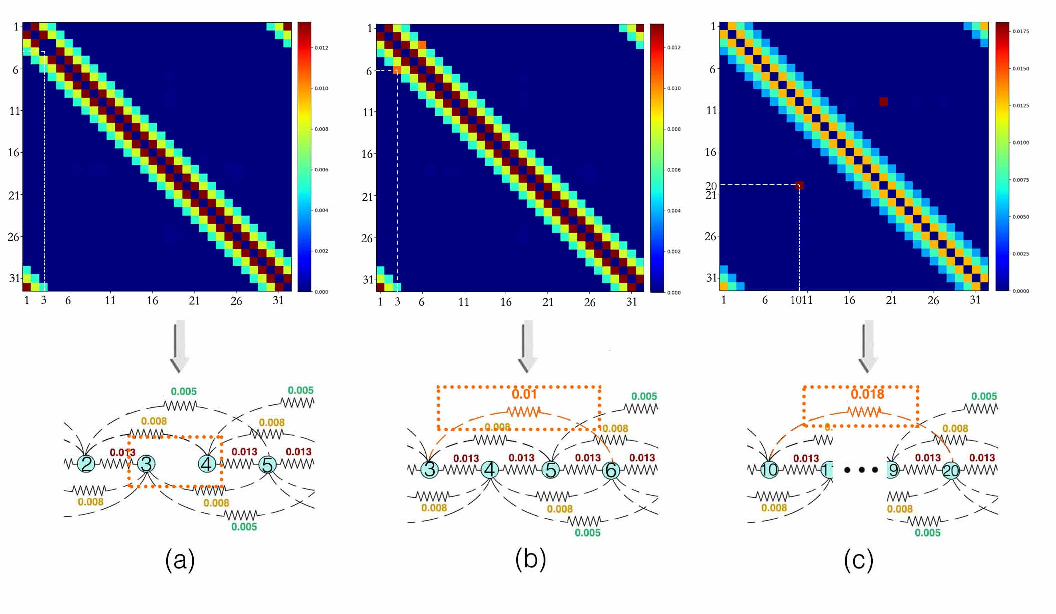}
\caption{Lattice structure anomaly detection:  (a) link breakage,  (b) differences in link strength, and (c) the presence of redundant links.  For clarity, the values on the diagonal are removed.}\label{fig_atten2}
\end{figure}

Fig. \ref{fig_atten2} depicts the detection of three situations in the system by GAHN. 
It can be inferred from the matrix in Fig. \ref{fig_atten2} (a) that the connection between the third and fourth particles is broken.
From Fig. \ref{fig_atten2} (b), it can be inferred that the interaction strength between the third and sixth particles is different from that between the third-order neighbors of other particles, which may be due to the presence of impurity defects or other reasons in the third and sixth particles. 
From Fig. \ref{fig_atten2} (c), it can be inferred that the tenth particle is different from other particles and has a long-range interaction with the twentieth particle.
These results are consistent with our hypothesis. 
This indicates that GAHN can infer the connectivity (or interaction) patterns of particles based on their trajectory changes. This ability has extremely high value in practical applications. For example, in the field of materials science, it is possible to optimize the performance of materials by controlling the formation of defects in particles. On the other hand, it is possible to detect whether materials have aging, damage, and other phenomena.

\subsection{Expand applications: Molecular chemical bond learning}
A molecule can be depicted as a graph $\mathcal{G} = (\mathcal{V}, \mathcal{E})$, where $\mathcal{V} = \{v_i\}_{i=1}^N$ represents the set of atoms constituting the molecule (nodes), and $\mathcal{E}$ represents the chemical bonds (edges) between these atoms. 
GAHN can be integrated with molecular dynamics research to determine atomic connectivity relationships (i.e. the presence of chemical bonds) through molecular motion trajectories.
We validated the capability of our model using two challenging molecular dynamics benchmark datasets: the MD17 dataset\cite{chmiela2017machine} and the MD22 dataset\cite{chmiela2023accurate}.

\begin{figure}[htbp]%
\centering
\includegraphics[scale=0.35]{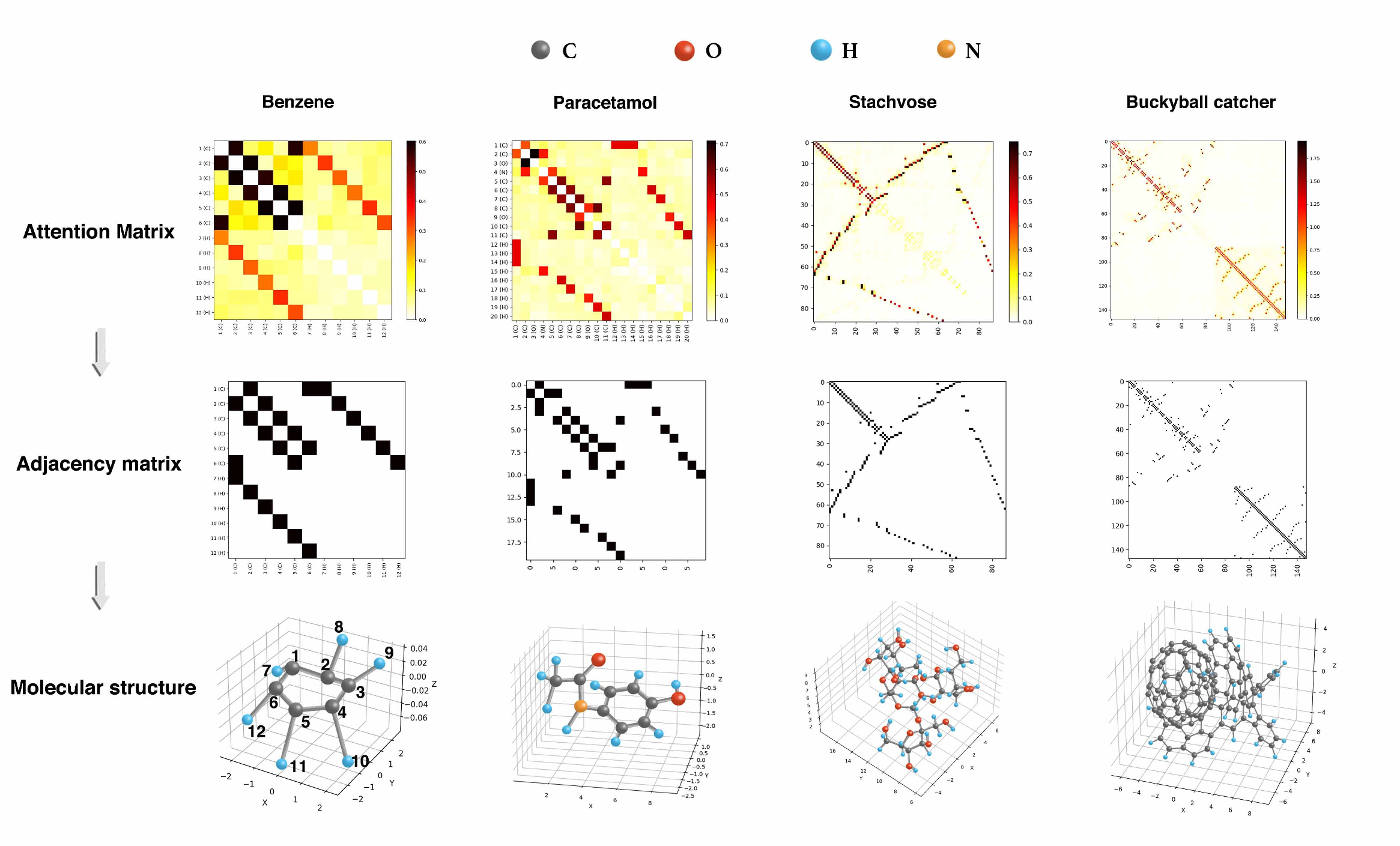}
\caption{Examples of small molecules. The three columns in the figure represent the average attention matrix $\bar{\mathbf{A}}$ of 10 training, the adjacency matrix obtained by setting a threshold for the $\bar{\mathbf{A}}$, and the chemical bond links obtained based on the adjacency matrix, respectively. The black area of the adjacency matrix represents a value of 1, indicating the existence of chemical bonds between the atoms corresponding to the rows and columns of the adjacency matrix. The coordinates of atoms in the molecule are randomly selected at a certain moment in the molecular trajectory. }
\label{fig_m_1}
\end{figure}

Fig. \ref{fig_m_1} shows some of the training results. In order to enhance visual clarity and intuitiveness, we also plotted the adjacency matrix results inferred from the attention matrix.
The first column of Figs. \ref{fig_m_1} shows the heatmaps of average attention matrix $\bar{\mathbf{A}}$ from 10 training. The second column displays the adjacency matrix $\mathbf{A}^o$, which is inferred from the processed average attention matrix $\bar{\mathbf{A}}$. Specifically, to extract the adjacency matrix, we apply a threshold, assigning a value of 0 to parts below the threshold and 1 to parts above. The threshold is set at 35\% of the maximum element in the average attention matrix.
The black area of the adjacency matrix represents a value of 1, indicating the existence of chemical bonds between the atoms corresponding to the rows and columns of the adjacency matrix.
The third column represents the chemical bond links inferred from the adjacency matrix, with atom coordinates randomly selected from a specific moment in the molecular trajectory. The size of an atom is directly proportional to its charge. Taking the benzene molecule in Fig. \ref{fig_m_1} as an example, the rows and columns of $\bar{\mathbf{A}}$ and $\mathbf{A}$ are consistent with the atomic labels in the third column of the molecular structure.
Because there is an element 1 between 12th row and 6th column of $\mathbf{A}$, this indicates the presence of a chemical bond between H$_{[12]}$ and C$_{[6]}$ in the molecular structure, where H$_{[12]}$ represents hydrogen with label 12.  
It can be seen that the learned chemical bonds are consistent with the reference chemical bonds provided by some current chemistry textbooks and research literature.  This helps to explore the chemical bond structures of unknown molecules in the future and speculate on their properties.
For more training details and molecular chemical bond learning results, please refer to the supplementary materials.

\begin{figure}[htbp]%
\centering
\includegraphics[scale=0.35]{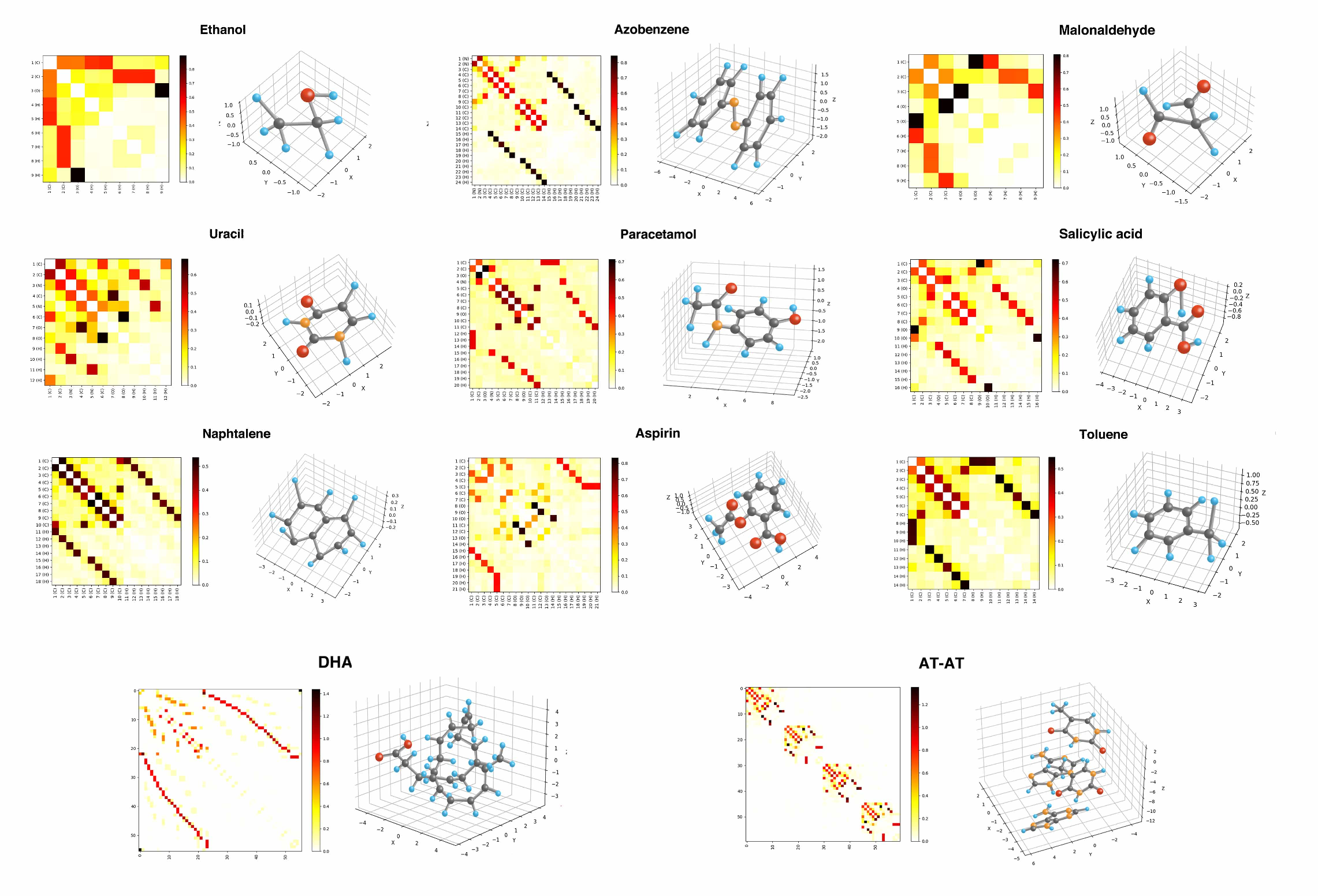}
\caption{The average attention matrix and the chemical bond connections inferred from it. }
\label{fig_ma}
\end{figure}

For each molecular, we selected 10,000 samples as the training set and 2,000 samples as the validation set for early stopping, with patience set to 20 epochs. For the Buckyball catcher, we used 5,000 samples for training and the remaining samples for validation. The batch size for the Buckyball catcher, stachyose, AT-AT, and DHA was set to 32, while for the remaining molecules, it was set to 256. All models were trained for a total of 5,000 epochs using the Adam optimizer with a learning rate of 0.001.
Fig. \ref{fig_ma} shows the average attention matrix of more tested molecules and the inferred chemical bond connections from it. 

\subsection{Compared to the classic GAT} 
\begin{figure}%
\centering
\includegraphics[scale=0.25]{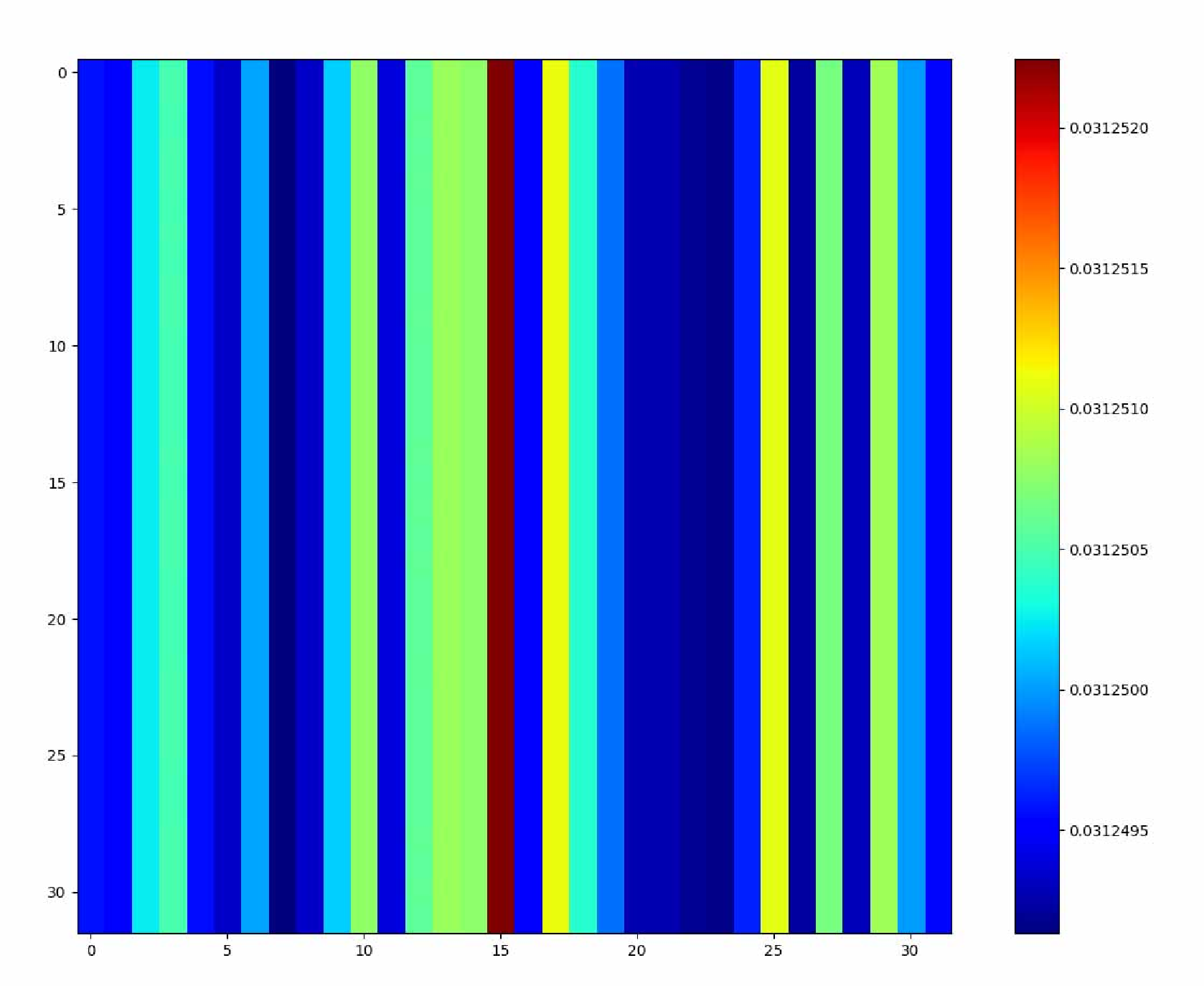}
\caption{The heatmap of the GAT coefficients defined in \cite{velivckovic2017graph} for KG-LRI.  }\label{fig_gatc}
\end{figure}

In this section, we compare our GAT construction with classical GAT \cite{velivckovic2017graph} methods by replacing the GAT coefficient $a_{ij}$ in our model with the GAT coefficient defined in \cite{velivckovic2017graph}. Taking KG-LRI as an example, Fig. \ref{fig_gatc} shows the output GAT coefficients, where we used a GAT layer defined in \cite{velivckovic2017graph} with 60 hidden layers. It can be seen that in the absence of a given system's real interaction, classical GAT coefficients are difficult to determine the true structure of the system.

\section{Conclusion}
In this work, we introduce GAHN, a neural network model that can capture the interactions and dynamic behaviors among particles based on trajectory data. This provides a more comprehensive understanding of lattice systems. The attention matrix learned by GAHN contains rich information about the system structure, such as which particles interact with each other, the proportion of interactions between different particles, whether the interaction potential energy between particles is even symmetric, whether there are connection defects, whether there are impurity defects causing different interactions, and so on.
In addition, GAHN can also predict the trajectory of the system. We benchmarked it with MLP, HNN, and  SympNet models and recorded the root mean square errors of their predicted trajectories and system energy. The results indicate that GAHN outperforms the baseline models in all cases, confirming the robustness and adaptability of our method.
GAHN offers the ability to provide graph structure for graph neural network-based methods, which compensates for the lack of prior structural knowledge required by graph neural network methods compared to conventional network methods.

GAHN has the potential to be widely applicable in scientific work. 
For example, in materials science, it can be used to detect defects in material connections and aging sites of materials.
In addition, it can integrate with other components to infer the appropriate link structure required for specific applications.
We validated it on a molecular dynamics dataset, successfully deducing the connectivity of molecular chemical bonds.
 The potential integration of GAHN with other machine learning paradigms also remains an exciting prospect for enhancing predictive performance and addressing a broader spectrum of scientific problems.

\section*{Funding data}
Y.X. Gao was  supported by the Science and Technology Development Plan Project of Jilin Province 20240101006JJ and NSFC grants [Grant Nos. 12371187, 12071065].
 H.-K. Zhang was partially supported by Simons Foundation Collaboration Grants for Mathematicians [Grant No. 706383].

\section*{Declaration of competing interest}
 The authors declare that they have no known competing financial interests or personal relationships that could have appeared to influence the work reported in this paper.



\bibliographystyle{elsarticle-num.bst}
\bibliography{refs}



%
%
%

\end{document}